\documentstyle[psfig,times]{mn}

\def\etal{et~al.}
\def\spose#1{\hbox to 0pt{#1\hss}}
\def\lta{\mathrel{\spose{\lower 3pt\hbox{$\mathchar"218$}}
     \raise 2.0pt\hbox{$\mathchar"13C$}}}
\def\gta{\mathrel{\spose{\lower 3pt\hbox{$\mathchar"218$}}
     \raise 2.0pt\hbox{$\mathchar"13E$}}}

\title[]{AGN--controlled cooling in elliptical galaxies} 

\author[P.~N.~Best \etal]{P.~N.~Best,$^1$\thanks{Email:
pnb@roe.ac.uk} C.~R.~Kaiser,$^2$ T. M. Heckman,$^3$ G. Kauffmann,$^4$ \\
$^1$ Institute for Astronomy, Royal Observatory Edinburgh, Blackford Hill,
Edinburgh EH9 3HJ \\
$^2$ School of Physics \& Astronomy, University of Southampton, Southampton
SO17 1BJ \\
$^3$ Department of Physics \& Astronomy, The Johns Hopkins University,
Baltimore, MD 21218, USA\\
$^4$ Max-Planck-Institut f{\"u}r Astrophysik, Karl-Schwarzschild-Str. 1,
D-85748 Garching, Germany
}

\begin{document}

\pagerange{\pageref{firstpage}--\pageref{lastpage}}
\pubyear{2005}

\label{firstpage}

\maketitle

\begin{abstract}
\noindent 
A long-standing problem for models of galaxy formation has been the
mismatch between the predicted shape of the mass function of dark matter
halos and the observed shape of the luminosity function of galaxies. The
number of massive halos is predicted to decrease as a power law ($N
\propto M^{-2}$) out to very large masses, while the galaxy luminosity
function cuts off exponentially at luminosities above $L_*$.  This implies
that the efficiency with which gas cools onto massive systems is lower
than expected. This letter investigates the role of radio--loud active
galactic nuclei (AGN) in continually re-heating the cooling gas. By
combining two observational results, the time--averaged energy output
associated with recurrent radio source activity is determined, as a
function of the black hole mass of the host galaxy: $\bar{H} = 10^{21.4}
(M_{\rm BH} / M_{\odot})^{1.6}$ W. It is shown that for massive elliptical
galaxies this radio--source heating balances the radiative energy losses
from the hot gas surrounding the galaxy. The recurrent radio--loud AGN
activity may therefore provide a self-regulating feedback mechanism
capable of controlling the rate of growth of galaxies.
\end{abstract}

\begin{keywords}
galaxies: active --- galaxies: evolution --- galaxies: stellar content ---
accretion --- radio continuum: galaxies
\end{keywords}

\section{Introduction}

In the widely accepted hierarchical clustering models of galaxy formation,
the growth of structure is controlled by the gravitational collapse of
haloes of dark matter, which merge together and accrete more mass to form
progressively larger structures. These models have provided an excellent
framework for describing the distribution of mass in the Universe. The
baryonic material, out of which stars and galaxies are formed, is
superimposed upon this dark matter distribution. As gas falls into the
gravitational potential well of dark matter haloes, it is initially
shock--heated to the virial temperature of the halo. This energy is then
radiated away, and the gas cools and condenses, and begins to form stars.
The properties of the resultant galaxies can be determined using
semi--analytic models (e.g. White \& Frenk 1991)\nocite{whi91}; these
adopt simple prescriptions for on--going physical processes, such as
cooling of the gas, star formation, chemical enrichment, merging of
galaxies, and feedback mechanisms, in order to predict the properties and
distribution of galaxies.

The galaxy luminosity function offers a powerful probe of the process of
galaxy formation. In early, simple, semi--analytic models, it was found
that the predicted shape of the galaxy luminosity function differed
greatly from that which is observed: many fewer galaxies are observed than
were predicted at both high and low masses. It was soon discovered that
the problem of the lack of faint galaxies could be solved by including in
the model the heating effects of supernovae (e.g. White \& Frenk
1991)\nocite{whi91} and photoionisation by the cosmic background
\cite{efs92}, which both act to reduce star formation efficiency at the
low mass end. The solution to the deficit of bright galaxies was
originally proposed to be due to cooling inefficiencies in massive
galaxies, but these effects were subsequently shown to be insufficient. In
order to reconcile the semi--analytic models with observations, some
additional suppression of the gas cooling in massive galaxies was
required. The generation of semi--analytic models of the late 1990's
introduced this in an ad-hoc manner (e.g. by artificially switching off
cooling in the most massive haloes), and were able to provide a very
successful description of galaxy properties in the nearby and high
redshift Universe (e.g. Kauffmann \etal\ 1999, Somerville \& Primack 1999,
Cole \etal\ 2000).\nocite{kau99b,som99,col00} In order to properly
understand galaxy formation, however, it is clearly important to
understand the physical processes involved in suppressing cooling in these
galaxies.

Active Galactic Nuclei (AGN) offer a promising solution to this problem
(e.g. Benson \etal\ 2003, Croton \etal\ 2005, Bower \etal\ 2005,
Scannapieco, Silk \& Bouwens 2005).\nocite{ben03,cro05,bow05,sca05} AGN
are predominantly found in the most massive galaxies (e.g. Kauffmann
\etal\ 2003)\nocite{kau03c}, and are sufficiently energetic that if their
output could be efficiently coupled to the gas it would be more than
sufficient to re-heat the gas and suppress further cooling. Prescriptions
for AGN feedback have recently been introduced into semi--analytic models
of galaxy formation, and both Croton \etal\ \shortcite{cro05} and Bower
\etal\ \shortcite{bow05} show that this feedback is able to solve the
issue of the bright end of the luminosity function, whilst simultaneously
solving other problems of galaxy formation models such as why the most
massive galaxies are so red.  These prescriptions are necessarily simple
in nature, however, and are not well grounded in observation. Indeed, the
form of the AGN feedback adopted is very different in the two
prescriptions (despite both providing good solutions) indicating that
there is still much work to be done in understanding exactly how and when
AGN feedback is important.

The focus of this letter is to determine the AGN feedback due to
radio--loud AGN. These are of particular interest because the expanding
radio source provides a direct way for the AGN output to be coupled to its
environment. Indeed, in clusters of galaxies containing powerful radio
sources, X--ray observations have revealed bubbles and cavities in the hot
intracluster medium, evacuated by the expanding radio source
(e.g. B\"ohringer \etal\ 1993, Carilli \etal\ 1994, McNamara \etal\ 2000,
Fabian \etal\ 2003).\nocite{boh93,car94b,mcn00,fab03} The (pV) energy
associated with these bubbles (which in turn arises from the mechanical
energy output of the radio jet) can be sufficient to balance the cooling
losses of some clusters, at least for a short period of time (e.g. Fabian
\etal\ 2003; B{\^ i}rzan \etal\ 2004).\nocite{fab03,bir04} Radio--loud AGN
have relatively short lifetimes ($10^7 - 10^8$ years), however, and so the
balancing of cooling will not be long--lived. Nonetheless, Best \etal\
\shortcite{bes05b} recently showed that over 25\% of the most massive
galaxies show radio--loud AGN activity, which implies that AGN activity
must be continually re--triggered.

In this letter, recent observational results are combined to show that the
time--averaged effect of heating by recurrent radio source activity
balances the radiative energy losses in the hot envelopes of the
ellipticals, across a wide range of black hole masses.
Section~\ref{method} provides a description of the observational results
upon which this work is built. In Section~\ref{results} the results are
discussed, the observations are compared with the prescriptions used in
the semi--analytic models, and conclusions are drawn. Throughout this
work, the cosmological parameters are assumed to have values of $\Omega_m
= 0.3$, $\Omega_{\Lambda} = 0.7$, and $H_0 = 70$\,km\,s$^{-1}$Mpc$^{-1}$.

\section{Methodology}
\label{method}

\subsection{The black-hole mass dependent radio luminosity function}
\label{meth_best}

Best \etal\ \shortcite{bes05a} constructed a sample of 2215 radio--loud
AGN with redshifts $0.03 < z < 0.3$, from the second data release (DR2) of
the Sloan Digital Sky Survey (SDSS), by comparing this with a combination
of the National Radio Astronomy Observatory (NRAO) Very Large Array (VLA)
Sky Survey (NVSS; Condon \etal\ 1998)\nocite{con98} and the Faint Images
of the Radio Sky at Twenty centimetres (FIRST) survey \cite{bec95}.  Best
\etal\ \shortcite{bes05b} compared these AGN with the radio--quiet
galaxies from the parent SDSS catalogue to derive the fraction of galaxies
that are radio--loud as a function of black hole mass ($M_{\rm BH}$, as
determined from the host galaxy velocity dispersion; e.g. Tremaine \etal\
2002)\nocite{tre02} and radio luminosity separately. They found that the
fraction of galaxies that are radio--loud depends very strongly on black
hole mass, going as $f_{\rm radio-loud} \propto M_{\rm BH}^{1.6}$. On the
other hand, the distribution of radio luminosities was found to be
independent of black hole mass. Best \etal\ were able to parameterise this
result using a mass--dependent broken power law model, such that overall
the fraction of sources that are radio--loud AGN brighter than some
1.4\,GHz radio luminosity $L$ is given by:

\begin{equation}
\label{eq1}
f_{\rm radio-loud} = f_0 \left(\frac{M_{\rm BH}}{10^8
M_{\odot}}\right)^{\alpha} \left[\left(\frac{L}{L_*}\right)^{\beta} +
\left(\frac{L}{L_*}\right)^{\gamma}\right]^{-1}
\end{equation}

\noindent with best--fit parameters of $f_0 = (3.5 \pm 0.4) \times
10^{-3}$, $\alpha = 1.6 \pm 0.1$, $\beta = 0.37 \pm 0.03$, $\gamma = 1.79
\pm 0.14$, $L_* = (3.2 \pm 0.5) \times 10^{24}$W\,Hz$^{-1}$. 

In a model in which radio source activity is continually re-triggered,
this result can instead be interpreted (probabilistically) as the {\it
fraction of its time} that a galaxy of given black hole mass spends as a
radio source brighter than a given radio luminosity.

\subsection{Radio luminosity {\it vs} mechanical energy output}
\label{meth_birzan}

\begin{figure}
\centerline{
\psfig{file=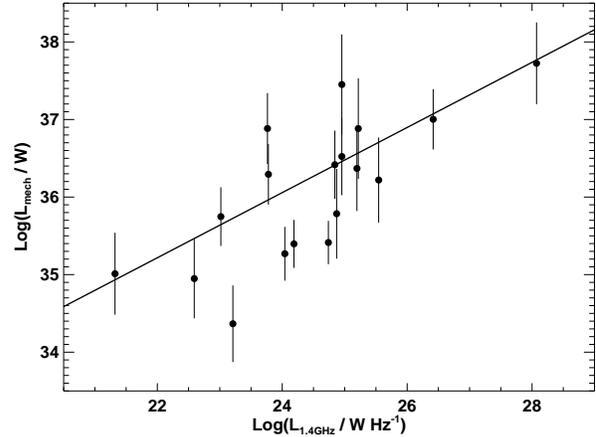,angle=90,width=8.6cm,clip=}
}
\caption{\label{lradlmechfig} The mechanical luminosity of radio sources,
as estimated by B{\^ i}rzan \etal\ (2004) from cavities and bubbles in the
X--ray gas, versus the monochromatic 1.4\,GHz radio luminosity. The solid
line shows an error--weighted best--fit to the relation (in linear
space). The errors on the mechanical luminosities come from different
assumptions in deriving the ages of the bubbles (see B{\^ i}rzan \etal\
for details).} 
\end{figure}

Monochromatic radio luminosity does not provide a good indicator of the
mechanical energy output of a radio source. Bicknell \shortcite{bic95}
estimate that the kinetic energy output of a radio jet is typically a
factor of 100--1000 higher than the total radio luminosity of a radio
source. In order to use the above results to determine the time averaged
heating output of radio--loud AGN, it is therefore necessary to derive a
conversion between radio luminosity and mechanical energy output. 

B{\^ i}rzan \etal\ \shortcite{bir04} studied the cavities and bubbles that
are produced in clusters and groups of galaxies due to interactions
between radio sources and the surrounding hot gas. They derived the (pV)
energy associated with the cavities; combining this with an estimate of
the age of the cavities provides an estimate of the mechanical luminosity
associated with the radio source. Figure~\ref{lradlmechfig} shows the
mechanical luminosities derived by B{\^ i}rzan \etal\ (summed over all
cavities associated with a given cluster or group) plotted against the
monochromatic 1.4\,GHz radio luminosity of the associated radio
sources. As B{\^ i}rzan \etal\ point out, there is no single scaling
factor between radio luminosity and mechanical luminosity. Nonetheless, it
is possible to obtain a reasonable fit to the data using a simple
power-law relation (shown as the solid line in Figure~\ref{lradlmechfig}):

\begin{equation}
\label{eq2}
\frac{L_{\rm mech}}{10^{36}{\rm W}} = (3.0 \pm 0.2) \left(\frac{L_{\rm
1.4GHz}}{10^{25}{\rm W Hz}^{-1}}\right)^{0.40 \pm 0.13}
\end{equation}

\noindent where the errors are calculated using bootstrap analysis.

The scatter in Figure~\ref{lradlmechfig} is larger than the error bars,
which suggests that there must be some additional dependencies in the
conversion from radio to mechanical luminosity. This is not surprising
since even for a source of fixed jet kinetic power the radio luminosity
changes as the source ages (e.g. Kaiser \etal\ 1997)\nocite{kai97b}.
Importantly, however, the offset from the best--fit line in
Figure~\ref{lradlmechfig} does not correlate with other properties of the
host galaxy (optical luminosity, velocity dispersion, environment)
indicating that for all host galaxies, Eq.~\ref{eq2} provides a reliable
measure of the mean mechanical luminosity associated with a source of
given radio luminosity.

Combining Eqs.~\ref{eq1} and~\ref{eq2} then gives the (probabilistic)
fraction of time that a galaxy of given black hole mass spends producing a
radio source of given mechanical luminosity, and from this the
time--averaged mechanical energy output of a galaxy can be derived as a
function of its black hole mass\footnote{Note that the conversion from
radio luminosity to mechanical luminosity given in Eq.~\ref{eq2} is
derived from radio sources in groups and clusters. It has been argued
(e.g. Barthel \& Arnaud 1996)\nocite{bar96a} that confinement of the radio
lobes by the higher pressure intracluster medium leads to boosted radio
luminosities compared to field radio sources of the same jet kinetic
power. This might imply that Eq.~\ref{eq2} would underestimate the
mechanical luminosities for radio sources outwith clusters. However, the
difference between cluster and field is only important for those sources
whose radio emission extends beyond their host galaxy, and hence the
extended surroundings control the pressure of the radio lobes. In
contrast, the radio sources which give rise to the bulk of radio source
heating are low luminosity sources (cf. Fig.~\ref{lradlheatfig}), which
tend to be compact and more confined to the host galaxy. The radio lobe
pressures are then more comparable to those of radio sources in clusters,
meaning that Eq.~\ref{eq2} will be approximately valid. Note also that
Eq.~\ref{eq2} gives values similar to those predicted by Bicknell
\shortcite{bic95}.}.  This evaluates to $\bar{H} = 10^{21.4} (M_{\rm BH} /
M_{\odot})^{1.6}$ W. The black hole mass dependence of this heating rate
arises entirely from the variation of the radio--loud fraction with black
hole mass, whilst the radio luminosity to mechanical luminosity conversion
determines the normalisation of the relation.

\subsection{Gas cooling rates in elliptical galaxy haloes}
\label{meth_osullivan}

The radiated energy losses from the haloes of hot gas surrounding
elliptical galaxies can be determined using X--ray observations.
O'Sullivan \etal\ \shortcite{osu01} have compiled a (not complete, but
relatively unbiased) catalogue of 401 early--type galaxies, for which they
have tabulated the X--ray luminosities ($L_X$, in erg\,s$^{-1}$), the
optical B--band luminosities ($L_B$, in units of the solar luminosity in
the B-band, $L_{B\odot} = 5.2 \times 10^{32}$erg\,s$^{-1}$), and the
T-type morphology parameter \cite{vau76}. The X--ray luminosities were
derived using a common spectral model and distance scale, and converted to
a pseudo--bolometric band of 0.088 to 17.25keV; O'Sullivan \etal\ show
that extending this energy band further makes $< 10$\% changes to the
bolometric X--ray luminosity.

The X--ray luminosity arises from two separate components of the galaxies:
the sum of discrete sources within the galaxy (X--ray binaries, discrete
stars, globular clusters, etc) and the cooling of the hot gas. The
luminosity of the discrete sources, being stellar in origin, should scale
roughly with the optical luminosity of the galaxy, and O'Sullivan~\etal\
estimate that this scales roughly as $L_X \approx 10^{29.5} L_B$. This can
dominate in low mass galaxies. For higher mass galaxies, the X--ray
luminosity is dominated by the emission from the hot gaseous haloes, and
O'Sullivan~\etal\ find that $L_X \propto L_B^2$ (see also the review by
Mathews \& Brighenti 2003).\nocite{mat03b}

\begin{figure}
\centerline{
\psfig{file=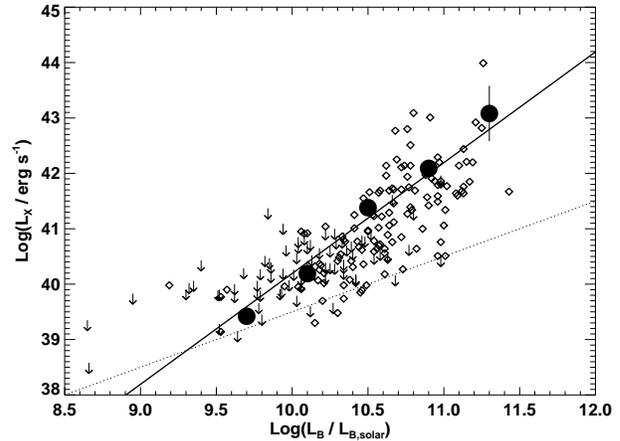,angle=90,width=8.6cm,clip=}
}
\caption{\label{lxlbfig} A plot of $L_X$ versus $L_B$ for the elliptical
galaxies in the sample of O'Sullivan~\etal\ (2001). The large filled
circles show the mean values of $L_X$ for galaxies in 5 bins of $L_B$
(each of width 0.4 in log$L_B$); the mean values were calculated from both
the observed values and upper limits, using the Kaplan--Meier estimator
within the {\sc ASURV} survival analysis package (LaValley \etal\ 1992).
The dotted line gives the expected contribution towards the X--ray
luminosities from the sum of discrete X-ray sources within the
galaxies. The solid line shows the predicted X--ray emission from the hot
haloes of the ellipticals, assuming that heating from radio--loud AGN
balances the cooling.}
\end{figure}

For the current paper, analysis is restricted to those sources with $T \le
-4$, that is, bona-fide elliptical galaxies. A plot of $L_X$ versus $L_B$
for these galaxies is shown in Figure~\ref{lxlbfig}; the dotted line on
that plot represents the average contribution to the X--ray luminosities
from the discrete sources. The residual effectively measures the cooling
energy losses from the haloes of the ellipticals, as a function of host
galaxy luminosity.

\subsection{Black hole mass versus bulge luminosity relation} 
\label{meth_mclure}

In order to compare the average radio--AGN heating rates (as a function of
black hole mass) with the average gas cooling rates (as a function of
B--band luminosity), the black hole masses are converted to galaxy
luminosities using the relation of McLure \& Dunlop \shortcite{mcl02}:
$\rm{log}(M_{\rm BH} / M_{\odot}) = -0.50 M_{\rm R} - 2.96$, where $M_R$
is the absolute R--band magnitude of the galaxy bulge. Because the sample
is restricted strictly to elliptical galaxies, the galaxy luminosity and
the bulge luminosity are identical. The R--band magnitude is converted to
a B-band luminosity using the average B--R rest--frame colour for
elliptical galaxies, B--R$\approx$1.2 (e.g. Blanton \etal\ 2003),
\nocite{bla03} and the absolute magnitude of the sun ($M_{{\rm B}\odot} =
5.48$).

\nocite{lav92} 

\section{Results and Discussion}
\label{results}

Figure~~\ref{lxlbfig} shows the $L_X$ versus $L_B$ relation for elliptical
galaxies from O'Sullivan \etal\ \shortcite{osu01}. The solid line on this
plot shows the time--averaged energy input into a galaxy by its recurrent
radio source activity. The agreement between these heating and cooling
rates across a wide range of host galaxy luminosities (masses) is
remarkable. Averaged over time, the energy supplied to a galaxy by its
recurrent radio source activity can balance the radiative energy losses
from its hot halo; thus, cooling of the gas is suppressed, and the radio
source activity may control the rate of growth of the elliptical galaxy.

\begin{figure}
\centerline{
\psfig{file=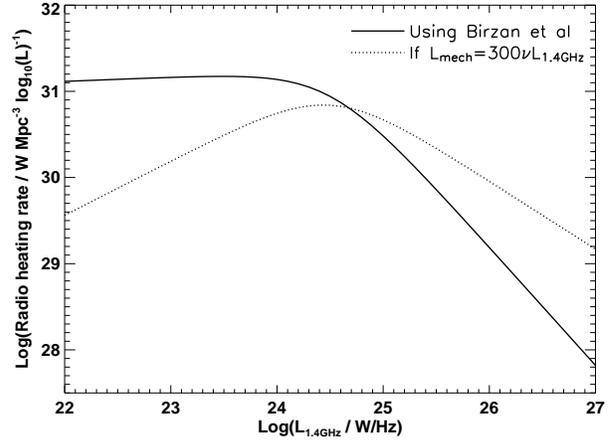,angle=90,width=8.6cm,clip=}
}
\caption{\label{lradlheatfig} The global rate of heating in the nearby
Universe produced by the sum total of all radio sources, as a function of
their radio luminosity. The solid line shows the result obtained in this
paper, deriving the mechanical luminosities of the sources from the fit to
the data in Figure~\ref{lradlmechfig}. The dashed line shows the very
different result that would be obtained assuming that the mechanical
luminosity were directly proportional to the radio luminosity (the line is
for $L_{\rm mech} = 300 \nu L_{1.4GHz}$, where the factor of 300 is
roughly that estimated by Bicknell 1995).}
\end{figure}

For this radio--source feedback model to work, it is necessary that the
majority of the mechanical energy supplied by the radio activity is
dissipated within the host galaxy halo. However, the most powerful radio
sources can have linear sizes very much larger than their host
galaxies. Combining the local radio luminosity function for AGN (e.g. Best
\etal\ 2005b)\nocite{bes05a} with the conversion from radio to mechanical
luminosity (Eq.~\ref{eq2}), the contribution to the global mechanical
heating rate from sources of different radio luminosities can be derived:
this is shown in Figure~\ref{lradlheatfig}. It is reassuring that the
heating is dominated by low luminosity sources, $L_{\rm 1.4GHz} \lta
10^{24}$W\,Hz$^{-1}$; such low luminosity sources tend to be small, and
their mechanical energy will indeed be confined predominantly to
size--scales of the host galaxy halo. These low luminosity sources are
also `on' for a significant fraction of the time (up to 30\% in the most
massive galaxies; see Best \etal\ 2005a),\nocite{bes05b} meaning that the
heat supply is reasonably continuous, as opposed to being in the
occasional short--lived explosive events associated with the (rarer) more
powerful sources.

The volume--averaged mechanical energy heating rate due to all radio
sources in the nearby Universe can be determined by integrating
Figure~\ref{lradlheatfig}; because of the flatness of the relation at low
radio luminosities, the total heating budget due to radio sources in the
nearby Universe depends upon the radio luminosity down to which the radio
luminosity function of AGN extends without change of slope\footnote{Note
that this result is very different from what would be obtained using the
simple assumption that the mechanical energy scales linearly with the
radio luminosity, when the bulk of the heating would arise from just a
small range of source luminosities near the break of the radio luminosity
function (cf. the dotted line in Fig.~\ref{lradlheatfig}), and the range
of radio luminosities considered would be less critical.}. Adopting
$L_{\rm 1.4GHz} = 10^{22}$W\,Hz$^{-1}$ as the lower limit gives a
zero--redshift volume--averaged heating rate of $\bar{H} = 4.0 \times
10^{31}$W\,Mpc$^{-3}$, but this value would nearly double if the slope of
the radio luminosity function for AGN remains unchanged down to
$10^{18}$W\,Hz$^{-1}$.

Regardless of the precise value, the observed local heating rate due to
radio--loud AGN is a factor of 10--20 lower than the local density of
radio--mode heating in the Croton \etal\ \shortcite{cro05} semi--analytic
model ($9 \times 10^{32}$W\,Mpc$^{-3}$, as calculated from their Eq.~10
and Fig.~3). Part of this difference is that the model cooling rates seem
to be over-estimated by a factor of 2--3, due to the schematic nature of
the underlying cooling flow model (White, Croton, private communication).
The remainder of the difference arises because the level of radio heating
in the Croton \etal\ model is tuned to balance cooling in {\it all}
massive galaxies, including the special case of brightest cluster galaxies
at the centres of extreme cooling flows. The observations presented here,
however, suggest only that radio source heating balances cooling radiation
losses in `typical' elliptical galaxies; there is no evidence for the
additional heating that would be required to control cooling on cluster
scales.  An alternative heating source is therefore required to solve the
cluster cooling flow problem. This could still be associated with radio
sources, but only if there is an additional mode of powerful radio source
activity associated specifically with brightest cluster galaxies; this is
beyond the test of the current observations.

The balance between heating and cooling rates in typical elliptical
galaxies is likely to occur through a feedback--controlled episodic
behaviour of the AGN activity (cf. Binney \& Tabor 1995, Kaiser \& Binney
2003):\nocite{bin95,kai03} the radio source activity heats the surrounding
gas; this then cools over time, until eventually some gas begins to
condense down onto the central galaxy; a small proportion of that gas is
driven down to the central regions of the galaxy, fuelling a new cycle of
AGN activity; the resulting radio source re-heats the gas and cuts off the
cooling once again.

One interesting question is why, in this scenario, the radio--loud AGN
fraction should scale as $M_{\rm BH}^{1.6}$. Best \etal\
\shortcite{bes05b} found that the occurrence of low luminosity radio--loud
AGN activity was completely independent of that of optical
(emission--line) AGN activity, and therefore argued that radio--loud AGN
were fuelled by the accretion of hot gas (as opposed to the cold gas which
fuels optical AGN through a standard thin accretion disk). Croton \etal\
\shortcite{cro05} showed that the Bondi--Hoyle accretion rate of hot gas
is $\dot{m}_{\rm Bondi} \approx G \mu m_p k T M_{\rm BH} / \Lambda$ (their
Eq.~28), where $\mu m_p$ is the mean mass of particles in the gas, and
$\Lambda$ is the cooling function. For an isothermal gas, $T \propto
\sigma^2$, which from the black hole mass versus velocity dispersion
relation gives $T \propto M_{\rm BH}^{0.5}$. $\Lambda$ is relatively
independent of temperature (and hence black hole mass) at the temperature
of elliptical galaxy haloes, and therefore the Bondi accretion rate scales
roughly as $M_{\rm BH}^{1.5}$, remarkably similar to the scaling of the
radio--loud fraction: the higher proportion of radio--loud AGN in more
massive galaxies may therefore arise naturally from the higher accretion
rates. The physics underlying this accretion must be more complicated,
since the Bondi accretion rate is for continual infall of gas whilst the
mass--scaling of the radio--loud AGN fraction represents a scaling of the
fraction of time that a given black hole is active, but it is certainly
intriguing that the two scale with mass in the same way, and that this
mass scaling matches that required in the Croton \etal\ model to solve the
current problems in the semi--analytic models.

\nocite{bic95}

In conclusion, therefore, observations show that the time--averaged
mechanical energy output associated with radio--source activity in massive
galaxies is found to almost exactly balance the cooling radiation losses
from the hot haloes of those galaxies, across a wide range of galaxy
masses. This implies that the cooling of the gas may be feedback
controlled by the radio activity. This feedback will control the growth
rate of all galaxies in haloes in which a quasi--static cooling halo of
hot gas has been established; in practice this means all dark matter
haloes more massive than $2-3 \times 10^{11} M_{\odot}$ (e.g. Bower \etal\
2005, Croton \etal\ 2005). The mass--scaling of the mechanical energy
output due to radio source activity matches both that predicted if they
are fuelled by Bondi accretion of hot gas from the cooling haloes, and
that required for radio--AGN feedback to successfully solve the problem of
the over-growth of massive galaxies in semi--analytic models.

\section*{Acknowledgements} 

PNB would like to thank the Royal Society for generous financial support
through its University Research Fellowship scheme. The authors thank Simon
White and Darren Croton for illuminating discussions about the cooling and
heating rates in the semi--analytic models, Ed Pope for useful discussions
about the recurrent radio source activity, and the referee for helpful
comments.

\bibliography{pnb} 
\bibliographystyle{mn} 
\label{lastpage}

\end{document}